\documentclass[
    ,final            
  ]
  {aipproc}

\layoutstyle{6x9}

\usepackage{latexsym, amsmath, amssymb}

\newcommand{\pt}{\partial}

\newcommand{\be}{\begin{equation}}
\newcommand{\ee}{\end{equation}}
\renewcommand{\[}{\begin{equation}}
\renewcommand{\]}{\end{equation}}

\newcommand{\bea}{\begin{eqnarray}}
\newcommand{\eea}{\end{eqnarray}}
\newcommand{\bsea}{\begin{subeqnarray}}
\newcommand{\esea}{\end{subeqnarray}}
\renewcommand{\d}{\mathrm{d}}

\begin{document}

\title{Bouncing and Colliding Branes}


\author{Jean-Luc Lehners}{
  address={DAMTP, CMS, Wilberforce Road, CB3 0WA, Cambridge, UK}
}

\begin{abstract}
In a braneworld description of our universe, we must allow for the
possibility of having dynamical branes around the time of the big
bang. Some properties of such domain walls in motion are discussed
here, for example the ability of negative-tension domain walls to
bounce off spacetime singularities and the consequences for
cosmological perturbations. In this context, we will also review a
colliding branes solution of heterotic M-theory that has been
proposed as a model for early universe cosmology.

\end{abstract}

\maketitle

\section{Introduction}

The duality between M-theory and the $E_8 \times E_8$ heterotic
string theory \cite{HW1} provides us with a braneworld picture of
our universe, in which one spatial dimension is a line segment
whose two boundaries are 10-dimensional (10d) branes containing
gauge fields. If 6 spatial dimensions are compactified on a
Calabi-Yau (CY) manifold, then we obtain a 5d picture, known as
heterotic M-theory, of a bulk spacetime bounded by two
(3+1)-dimensional boundary branes \cite{LOSW1,LOSW2}. The action
is given by
 \be
S = \int_{5d} \sqrt{-g} \,[R - \frac{1}{2}(\pt \phi)^2 - 6\alpha^2
e^{-2 \phi}] \pm 12 \alpha \int_{4d,\,y=\mp1} \sqrt{-g} e^{-
\phi}, \label{Action5d} \ee where the scalar field $\phi$
parameterises the volume of the CY manifold, and $\alpha$ the flux
on it. Since the boundary branes contain Standard Model-type gauge
fields, we identify one of these branes with our currently visible
universe. If we take this braneworld picture seriously, then we
know from the constancy of the coupling constants that the
distance between the branes, as well as the size of the internal
CY manifold, must have been varying exceedingly slowly, if at all,
since shortly after the big bang. However, we must allow for the
possibility that the branes were truly dynamic around the time of
the big bang. Such dynamic branes can have important consequences
for cosmology: for example, the basis of the ekpyrotic and cyclic
models is the identification of the big bang with a collision of
the two boundary branes \cite{Khoury:2001wf,Steinhardt:2001st}.

Apart from a purely cosmological motivation, it is in any case of
theoretical interest to study time-dependent branes regarding
their behaviour with respect to singularities. In 5d the branes
are domain walls, and the geometry can therefore be described in
terms of a linear harmonic function $h(y),$ where $y$ denotes the
line segment coordinate. At the location of the branes, the
harmonic function contains kinks whose magnitudes are related to
the brane tensions. In a time-dependent setting, one can thus
imagine the slope as well as the height of this harmonic function
to be varying. Moreover, a zero of the harmonic function
corresponds to a timelike naked singularity. One might therefore
be worried that the negative-tension brane, at whose location the
harmonic function is always lowest, could crush into this
singularity \cite{Chen:2005jp}. A separate worry is that naively
one would expect a collision of the two boundary branes to be
accompanied by the internal manifold shrinking to zero size and
hence the couplings and curvature invariants to be blowing up
\cite{Lehners:2002tw}. Here we will show how both of these
potential catastrophes can be avoided, and how avoiding them can
lead to unexpected benefits.

\section{Colliding Branes - Bouncing Branes}

We are looking for a solution in which the collision of the
boundary branes is the least singular possible. Thus we impose the
boundary conditions that at the collision, the scale factors on
the branes and the CY size approach a finite and non-zero constant
\cite{Lehners:2006pu}. Then the spacetime tends to compactified
Milne $\times \mathbb{R}^3$ (plus a static CY) at the collision,
which means that curvature invariants are small close to the
collision. Hence all eventual higher-derivative corrections will
also be small, and we can trust our solution right up to the
collision (As an aside, note that the brane tensions go to zero at
the collision. Also, close to the collision, one expects all
relevant modes to be winding modes, which are well-behaved at
small orbifold size, see \cite{Turok:2004gb} for details.). It
turns out that imposing the boundary conditions enunciated above
determines the solution almost uniquely \cite{Lehners:2006pu}:
there is just one free parameter, namely the velocity of the
branes at the collision.

\begin{figure}
  \includegraphics[height=.25\textheight]{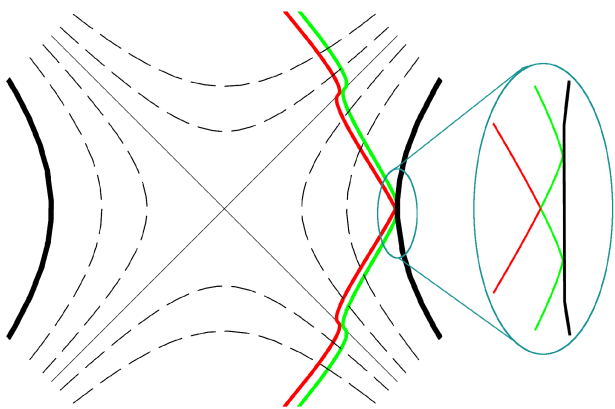}
  \caption{A Kruskal plot of
the colliding branes solution. The timelike naked singularity is
denoted by thick black lines. Here the trajectory of the
positive-tension brane is shown in red and that of the
negative-tension brane in green.  The collision of the branes, as
well as the two bounces of the negative-tension brane off the
naked singularity, are shown at a magnified scale in the inset.}
\end{figure}

In describing time-dependent branes, there exist two coordinate
systems which are particularly useful. In the first, which has
been used implicitly up to now, the branes are kept at fixed
coordinate locations and the bulk is dynamical, while in the
second, the branes are moving in a static bulk spacetime. The
latter description turns out to be more convenient in presenting
the colliding branes solution. The bulk metric and scalar field
are given by \be \label{staticbulk2} \d s^2 = -(\alpha^2 r^2 - \mu
r^4)\,\d t^2 + (\alpha^2 r^2-\mu r^4)^{-1} \, r^{12} \, \d r^2 +
r^2 \d\vec{x}^2, \qquad e^\phi = r^6. \ee Here $r$ denotes the
coordinate transverse to the branes, and $\mu$ is related to the
collision velocity. At $r=0$ there is a timelike naked
singularity, which, after changing coordinate systems, would
correspond to a zero of the linear harmonic function. The branes'
motion in the static bulk is determined by solving their Israel
matching conditions. From Figure 1 it can be seen that the
negative-tension brane grazes the singularity twice. In the
absence of matter on the branes, the brane actually touches the
singularity, leading to a catastrophic crunch. However, if matter
is added to the branes, then the negative-tension brane can bounce
back before reaching $r=0$, thus shielding the positive-tension
brane as well as the bulk from the singularity
\cite{Lehners:2007nb}. The precise conditions for a bounce to
occur depend on the coupling of the brane-bound matter to the CY
volume scalar, and have been presented in \cite{Lehners:2007nb}.
Typically, for a given matter type, there is a continuous,
semi-infinite range of couplings that lead to a bounce. Also, for
one specific value of the coupling, only the brane trajectory is
altered, and the bulk remains unchanged.

\begin{figure}
  \includegraphics[height=.25\textheight]{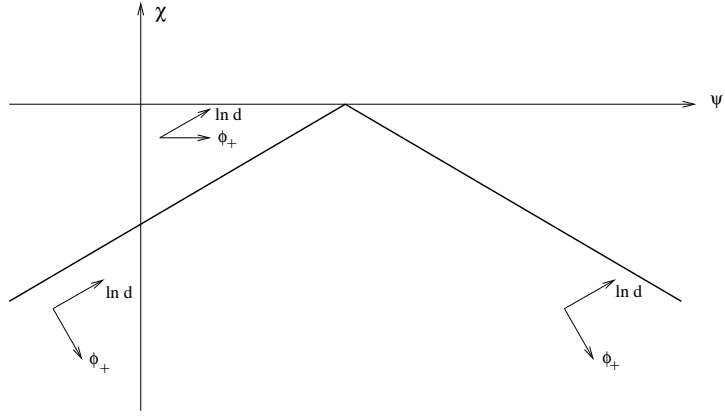}
  \caption{The trajectory of the colliding branes solution as seen in the $\psi$ - $\chi$ plane.
The brane collision occurs as $\psi$, $\chi \rightarrow - \infty$.
Also shown are the directions of increasing distance between the
boundary branes ($\ln{d}$) and increasing CY volume at the
location of the positive-tension brane ($\phi_+$).}
\end{figure}


It is instructive to see what the bouncing and colliding branes
described above look like from the 4d point of view. The
surprisingly simple 4d effective theory is given by \be
S_{\mathrm{moduli}} = 6 \int_{4d} [-\dot{a_4}^2 + a_4^2
(\dot{\psi}^2 + \dot{\chi}^2)], \label{ActionMSA2} \ee where $a_4$
is the effective 4d scale factor, and the two scalars $\psi$ and
$\chi$ are related to the inter-brane distance and the size of the
CY manifold \cite{Lehners:2006ir}. While the brane collision
corresponds to a big crunch/big bang ($a_4 \rightarrow 0$), it
turns out that in the $\psi-\chi$ plane, the bounce corresponds to
a reflection off a boundary of moduli space, see Figure 2. This
boundary can be inferred from the requirement of positivity of the
volume of the internal manifold, and it corresponds to imposing
$\chi \leq 0.$ The reflection has important consequences for
cosmological perturbations around the background solution
\cite{Lehners:2007ac}. Indeed, it is well-known that a
non-geodesic motion of the background trajectory has the effect of
converting entropy perturbations into curvature perturbations
\cite{Gordon:2000hv}. Here this conversion happens very rapidly,
and is thus spectrum-preserving. However, in order to have a
working model of early universe cosmology, we must ensure that we
can obtain the right spectrum of entropy perturbations in the
first place. A nearly scale-invariant spectrum is generated if we
add two negative nearly-exponential potentials for the scalar
fields $\psi$ and $\chi$ in the region where perturbations are
produced \cite{Lehners:2007ac} (It has so far been impossible to
directly compute the forces between the boundary branes, and so
the best we can do at the moment is to simply assume a certain
form for the scalar field potentials.). Then the background
trajectory falls off a ridge of the potential, and is thus
inherently unstable (This means that one will need a theory of
initial conditions. For a proposed solution to this problem in a
similar setting, see \cite{Buchbinder:2007tw}.). The reflection of
the background trajectory off $\chi=0$ then converts the entropy
perturbations into nearly scale-invariant curvature perturbations
shortly before the big crunch/big bang transition, and in the
ensuing expanding phase, the curvature perturbations grow.

\section{Conclusion}

Perhaps the main conclusion to be drawn from this work is that in
the well-motivated setting of heterotic M-theory, the
singularities threatening to plague a dynamical braneworld model
of the early universe can be dealt with consistently and can even
have unanticipated and beneficial side effects, such as the bounce
of the negative-tension brane, which not only stabilises the
braneworld, but at the same time solves the problem of how to
obtain growing-mode curvature perturbations from a contracting
phase.



\begin{theacknowledgments}

It is a pleasure to thank Paul McFadden, Paul Steinhardt and Neil
Turok for collaboration on this work. My research is supported by
PPARC.

\end{theacknowledgments}



\bibliographystyle{aipproc}   

\bibliography{BouncingColliding}

\begin{thebibliography}{14}
\expandafter\ifx\csname natexlab\endcsname\relax\def\natexlab#1{#1}\fi
\providecommand{\enquote}[1]{``#1''}
\expandafter\ifx\csname url\endcsname\relax
  \def\url#1{\texttt{#1}}\fi
\expandafter\ifx\csname urlprefix\endcsname\relax\def\urlprefix{URL }\fi
\providecommand{\eprint}[2][]{\url{#2}}

\bibitem[Horava and Witten(1996)]{HW1}
P.~Horava, and E.~Witten, \emph{Nucl. Phys.} \textbf{B460}, 506--524 (1996).

\bibitem[Lukas et~al.(1999{\natexlab{a}})]{LOSW1}
A.~Lukas, B.~A. Ovrut, K.~S. Stelle, and D.~Waldram, \emph{Phys. Rev.}
  \textbf{D59}, 086001 (1999{\natexlab{a}}).

\bibitem[Lukas et~al.(1999{\natexlab{b}})]{LOSW2}
A.~Lukas, B.~A. Ovrut, K.~S. Stelle, and D.~Waldram, \emph{Nucl. Phys.}
  \textbf{B552}, 246--290 (1999{\natexlab{b}}).

\bibitem[Khoury et~al.(2001)]{Khoury:2001wf}
J.~Khoury, B.~A. Ovrut, P.~J. Steinhardt, and N.~Turok, \emph{Phys. Rev.}
  \textbf{D64}, 123522 (2001).

\bibitem[Steinhardt and Turok(2002)]{Steinhardt:2001st}
P.~J. Steinhardt, and N.~Turok, \emph{Phys. Rev.} \textbf{D65}, 126003 (2002).

\bibitem[Chen et~al.(2006)]{Chen:2005jp}
W.~Chen, Z.~W. Chong, G.~W. Gibbons, H.~Lu, and C.~N. Pope, \emph{Nucl. Phys.}
  \textbf{B732}, 118--135 (2006).

\bibitem[Lehners and Stelle(2003)]{Lehners:2002tw}
J.-L. Lehners, and K.~S. Stelle, \emph{Nucl. Phys.} \textbf{B661}, 273--288
  (2003).

\bibitem[Lehners et~al.(2007{\natexlab{a}})]{Lehners:2006pu}
J.-L. Lehners, P.~McFadden, and N.~Turok, \emph{Phys. Rev.} \textbf{D75},
  103510 (2007{\natexlab{a}}).

\bibitem[Turok et~al.(2004)]{Turok:2004gb}
N.~Turok, M.~Perry, and P.~J. Steinhardt, \emph{Phys. Rev.} \textbf{D70},
  106004 (2004).

\bibitem[Lehners and Turok(2007)]{Lehners:2007nb}
J.-L. Lehners, and N.~Turok  (2007), \eprint{arXiv:0708.0743 [hep-th]}.

\bibitem[Lehners et~al.(2007{\natexlab{b}})]{Lehners:2006ir}
J.-L. Lehners, P.~McFadden, and N.~Turok, \emph{Phys. Rev.} \textbf{D76},
  023501 (2007{\natexlab{b}}).

\bibitem[Lehners et~al.(2007{\natexlab{c}})]{Lehners:2007ac}
J.-L. Lehners, P.~McFadden, N.~Turok, and P.~J. Steinhardt
  (2007{\natexlab{c}}), \eprint{hep-th/0702153}.

\bibitem[Gordon et~al.(2001)]{Gordon:2000hv}
C.~Gordon, D.~Wands, B.~A. Bassett, and R.~Maartens, \emph{Phys. Rev.}
  \textbf{D63}, 023506 (2001).

\bibitem[Buchbinder et~al.(2007)]{Buchbinder:2007tw}
E.~I. Buchbinder, J.~Khoury, and B.~A. Ovrut  (2007), \eprint{arXiv:0706.3903
  [hep-th]}.

\end{thebibliography}


\end{document}